\documentclass[9pt,conference]{IEEEtran}
\usepackage{subcaption}
\usepackage{pifont}
\usepackage{multirow}
\usepackage{makecell}

\usepackage{waspaa25} 

\usepackage{bm} 

\title{Long-Context Modeling Networks for Monaural Speech Enhancement: A Comparative Study}

\name{Qiquan Zhang$^{1,*}$,
      Moran Chen$^{2,*}$,
      Zeyang Song$^{3}$,
      Hexin Liu$^{4}$,
      Xiangyu Zhang$^{1}$,
      Haizhou Li$^{5,3}$\thanks{*Equal Contribution}}
\address{$^{1}$The University of New South Wales, Sydney, Australia \quad
$^{2}$Harbin Institute of Technology, Shenzhen, China \\
$^{3}$National University of Singapore, Singapore \quad
$^{4}$Nanyang Technological University, Singapore \\
$^{5}$Shenzhen Research Institute of Big Data, The Chinese University of Hong Kong, Shenzhen, China 
}

\begin{document}

\maketitle

\begin{abstract}
\textcolor{black}{Advanced long-context modeling backbone networks, such as Transformer, Conformer, and Mamba, have demonstrated state-of-the-art performance in speech enhancement. However, a systematic and comprehensive comparative study of these backbones within a unified speech enhancement framework remains lacking. In addition, xLSTM, a more recent and efficient variant of LSTM, has shown promising results in language modeling and as a general-purpose vision backbone. In this paper, we investigate the capability of xLSTM in speech enhancement, and conduct a comprehensive comparison and analysis of the Transformer, Conformer, Mamba, and xLSTM backbones within a unified framework, considering both causal and noncausal configurations. Overall, xLSTM and Mamba achieve better performance than Transformer and Conformer. Mamba demonstrates significantly superior training and inference efficiency, particularly for long speech inputs, whereas xLSTM suffers from the slowest processing speed.}

\end{abstract}

\section{Introduction}
\label{sec:intro}
Speech enhancement is the process of isolating clean speech in the presence of background noise, thereby improving speech quality and intelligibility. It is often employed as a front-end in various speech processing systems, such as speech recognition, speaker verification, mobile communication, and audio-video conferencing. Traditional unsupervised solutions~\cite{loizou2013,mmse2016,zhang2019,zhang2019fast} relying on statistical relations struggle to suppress rapidly varying background interference. Supervised learning approaches~\cite{wang2018supervised}, which leverage large-scale noisy-clean pairs to train deep neural networks (DNNs) to produce clean speech given noisy inputs, have recently dominated the field of speech enhancement.

Among these methods, common techniques involve spectral masking~\cite{wang2014training}, spectral mapping~\cite{xu2014regression}, and waveform-domain approaches~\cite{kolbaek2020loss,convtasnet}. Waveform-domain approaches~\cite{cleanunet,realse} often employ an encoder-masking-decoder network architecture to directly predict the clean waveform from the noisy waveform input. In spectral masking and mapping methods, given the noisy spectral representation, DNNs are optimized to estimate either a spectral mask (e.g., complex ideal ratio mask (cIRM)~\cite{williamson2015complex} and phase-sensitive mask (PSM)~\cite{erdogan2015phase}) or spectral representation of clean speech (e.g., magnitude spectrum, complex spectrum, and log-power spectrum). Generative approaches~\cite{lemercier2024diffusion,9632770,wavelet} follow a different paradigm by learning the probability distribution over clean speech data, enabling the generation of plausible clean speech conditioned on input noisy speech. More recent studies~\cite{storm,segmse} attempt to employ diffusion models for generative speech enhancement, demonstrating promising results.

To model long-context dependencies of speech signals, various network architectures have been explored for speech enhancement. Long short-term memory (LSTM) \textcolor{black}{Recurrent Neural Networks (RNNs)}~\cite{weninger2015speech} suffer from the inherent nature of sequential modeling. With the capability to capture long-context information in parallel, temporal convolution networks (TCNs) have shown comparable or better performance than LSTMs~\cite{tfaj,restcnsa,mbtcn}. TCNs struggle with limited context receptive fields built through stacked dilated 1-D convolution layers. The use of Transformer and its variants (e.g., Conformer) has led to cutting-edge results in a wide range of speech process tasks, including speech enhancement~\cite{sepformerstft,10446337,length}. The core component in Transformer is the self-attention mechanism that can dynamically model global dependencies of the entire speech sequence. The self-attention suffers from quadratic computational complexity with respect to the sequence length~\cite{ripple,tii}, which motivates exploring alternatives to Transformers. More recently, state space models (SSMs) have emerged as promising alternative network architecture~\cite{gu2023mamba}. Mamba, a state space model with a selection mechanism, adaptively adjusts the SSM parameters based on the input sequence and has yielded outstanding results in various speech applications~\cite{mamba,miyazaki2024exploring}, including speech enhancement~\cite{mamba}.

Meanwhile, the extended LSTM (xLSTM)~\cite{beck2024xlstm} introduces exponential gating with appropriate normalization and stabilization techniques while abandoning memory mixing, obtaining two variants, i.e., mLSTM and sLSTM. xLSTM has demonstrated performance comparable or better than Transformers and Mamba for language modeling and as a generic vision backbone~\cite{alkin2024vision}. State-of-the-art results in speech enhancement have been achieved by these backbones (Transformer, Conformer, and Mamba)~\cite{9746171,mpsenet,mambadc}, involving different design choices, such as model structure, loss function, and input feature (e.g., frame length and shift). A comprehensive and fair comparative study of these backbone networks in a unified speech enhancement framework remains missing. In this paper, we investigate the efficacy of xLSTM in speech enhancement and systematically compare Transformer, Conformer, Mamba, and xLSTM backbones within a unified framework, in terms of model performance, model size, training speed, and inference speed.


\textcolor{black}{This paper is organized as follows. We formulate the problem of neural time-frequency (T-F) speech enhancement in Section~\ref{sec:2}. We present the unified framework and long-context modeling backbones in Section~\ref{sec:3}. Section~\ref{sec:4} reports and analyzes the experimental results. Section~\ref{sec:5} concludes this paper.}

\section{Neural T-F Speech Enhancement}\label{sec:2}
\subsection{Problem Formulation}\label{sec:2.1}
Let $\bm{s}\in\mathbb{R}^{1\times N}$ and $\bm{d}\in\mathbb{R}^{1\times N}$ denote clean speech and noise waveforms, respectively, where $N$ indicates the number of time samples. The observed noisy mixture $\bm{x}\in\mathbb{R}^{1\times N}$ can be modeled as $\bm{x}=\bm{s}+\bm{d}$. The raw waveform representation is converted into a T-F representation using the short-time Fourier transform (STFT): $\mathbf{X}_{l,k}=\mathbf{S}_{l,k}+\mathbf{D}_{l,k}$, $\mathbf{S}_{l,k}$, $\mathbf{D}_{l,k}$, and $\mathbf{X}_{l,k}$, denoting the STFT coefficients of the clean speech, noise, and noisy mixture, respectively, at the $k$-th frequency bin of the $l$-th time frame. One common method in neural T-F speech enhancement involves employing a DNN to estimate a T-F mask $\mathbf{M}$ from the noisy STFT spectrum input: $\mathbf{M}=f_{\text{DNN}_{\theta}}(\mathbf{X})$. The mask is then applied to the noisy spectrum to obtain the clean spectrum estimate: $\widehat{\mathbf{S}}=\widehat{\mathbf{M}}\odot\mathbf{X}$. Here, we employ the commonly used PSM~\cite{erdogan2015phase}, formulated as

\begin{equation}
\mathbf{M}_{l,k} =
\frac{|\mathbf{S}_{l,k}|}{|\mathbf{X}_{l,k}|}
\cos\left(\operatorname{arg}(\mathbf{S}_{l,k}) - \operatorname{arg}(\mathbf{X}_{l,k})\right)
\end{equation}
where $|\cdot|$ extracts the magnitude, $\operatorname{arg}(\mathbf{S}_{l,k})$ and $\operatorname{arg}(\mathbf{X}_{l,k})$ represent the clean and noisy spectral phase, respectively.

\section{Network Architecture}\label{sec:3}

\begin{figure}[!ht]
\centering
\centerline{\includegraphics[width=0.5\columnwidth]{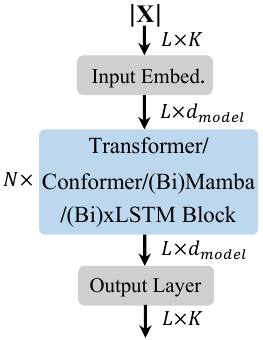}}
\caption{\textcolor{black}{Overall diagram of the unified (standard) neural T-F speech enhancement framework~\cite{tfaj,mhanet,10446337,deepmmse}.}}
\label{fig1}
\end{figure}

\textcolor{black}{To ensure a systematic and fair comparative study, we employ a unified neural speech enhancement framework to evaluate the long-context modeling backbone networks (Transformer, Conformer, Mamba, and xLSTM). To set the stage for this study, in Fig.~\ref{fig1}, we illustrate a typical neural solution to T-F speech enhancement~\cite{tfaj,mhanet,10446337,deepmmse}, which is referred to as the backbone network in this paper. Our comparative study involves four widely studied backbone networks: Transformer~\cite{tfaj,10446337}, Conformer~\cite{gulati2020conformer,tfaj}, Mamba~\cite{mamba,gu2023mamba}, and xLSTM~\cite{beck2024xlstm}.}

\textcolor{black}{Here, we describe the workflow of the network architecture. It takes as input the STFT spectral magnitude of noisy waveform $|\mathbf{X}|\in\mathbb{R}^{L\times K}$ of $L$ frames, each comprising $K$ frequency bins. The input $|\mathbf{X}|$ is first passed through an projection layer and projected into a latent embedding with a dimension of $d_{model}$. The embedding layer is a 1-D convolution layer pre-activated by a frame-wise layer normalization followed by the ReLU activation function. The input embedding is then passed through $N$ stacked building blocks, including Transformer, Conformer, Mamba, or xLSTM blocks for feature transformation. Following the stacked building blocks is the output layer, a 1-D convolution layer activated by a sigmoidal function, which generates the estimated mask.}

\textcolor{black}{\textbf{Backbone Networks.} \textcolor{black}{In our comparative study, we consider both causal (online applications) and non-causal configurations (offline applications). The standard self-attention module is non-causal. For causal Transformer and Conformer backbones, a causal attention mask is applied to mask out the upper triangular scores of the self-attention matrix to block the interactions from future frames~\cite{10446337}. In addition, recent findings~\cite{10446337} demonstrate that positional embedding (PE) is not quite helpful for Transformer speech enhancement in a causal configuration, whereas non-causal Transformer significantly benefits from the use of PE. To enable a systematic and comprehensive comparison, we introduce both absolute (Sinusoidal PE, i.e., SinPE)~\cite{transformer} and relative (Rotary PE, i.e., RoPE)~\cite{RoPE} position embeddings to further boost non-causal Transformer speech enhancement.}}

\begin{figure}[!htbp]
\centering
\includegraphics[width=0.55\columnwidth]{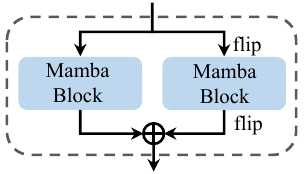}
\caption{\textcolor{black}{Illustration of the bidirectional Mamba (BiMamba) block~\cite{mamba}. $\oplus$ represents the element-wise addition.}}
\label{fig2}
\vspace{-1.0em}
\end{figure}

The original Mamba and xLSTM models are causal. For non-causal configuration, bidirectional Mamba (BiMamba) and bidirectional xLSTM (BixLSTM) are evaluated. Here, we employ the External BiMamba (illustrated in Figure~\ref{fig2}) proposed in~\cite[Figure~1(b)]{mamba}, which has been shown to slightly outperform the Inner BiMamba~\cite[Figure~1(a)]{mamba}. Each external BiMamba block consists of two parallel Mamba blocks, a forward block and a backward block.

\begin{figure}[!ht]
\vspace{-0.9em}
\centering
\begin{subfigure}[t]{0.43\columnwidth}
\centerline{\includegraphics[width=0.7\columnwidth]{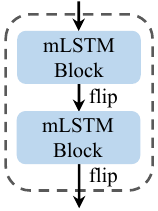}}
\caption{}
\label{fig3a}
\end{subfigure}
\begin{subfigure}[t]{0.55\columnwidth}
\centerline{\includegraphics[width=0.90\columnwidth]{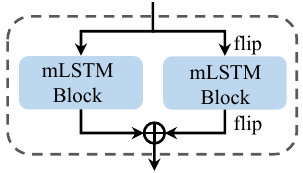}}
\caption{}
\label{fig3b}
\end{subfigure}
\caption{\textcolor{black}{Illustration of (a) the cascaded bidirectional xLSTM (C-BixLSTM) block~\cite{alkin2024vision} and (b) the parallel bidirectional xLSTM (P-BixLSTM) block.}}
\label{fig3}
\vspace{-0.5em}
\end{figure}
\begin{figure}[!htpb]
\centering
\includegraphics[width=0.95\columnwidth]{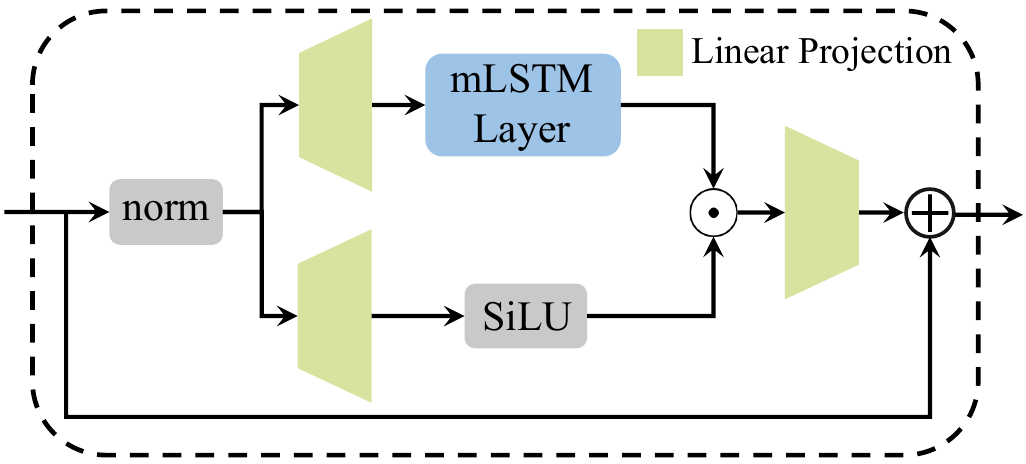}
\caption{\textcolor{black}{Illustration of the workflow of mLSTM block. $\odot$ and SiLU respectively represent the element-wise multiplication and activation function.}}
\label{fig4}
\vspace{-1.0em}
\end{figure}


xLSTM~\cite{beck2024xlstm} introduces two building blocks: sLSTM and mLSTM. For BixLSTM, following the structure design of Vision-LSTM vision backbone~\cite{alkin2024vision}, we investigate the cascaded BixLSTM (C-BixLSTM) for speech enhancement, using mLSTM as the building block. As shown in Figure~\ref{fig3}\,(a), the C-BixLSTM block consists of cascaded forward and backward xLSTM blocks. In addition, inspired by the success of BiMamba~\cite{mamba}, we also explore the same structure, i.e., parallel BixLSTM (P-BixLSTM) model (shown in Fig.~\ref{fig3}\,(b)), for speech enhancement. The P-BixLSTM block consists of two parallel xLSTM blocks, a forward block, and a backward block. The detailed diagram of the mLSTM block is illustrated in Fig.~\ref{fig4}. \textcolor{black}{We refer readers to the original publication~\cite{beck2024xlstm,alkin2024vision} for more details about xLSTM block.}

\section{Experiments} \label{sec:4}
\subsection{Datasets and Feature Extraction}
\textcolor{black}{The training data for clean speech comprises $28\,539$ utterances (approximately 100 hours) sourced from the LibriSpeech \textit{train-clean-100} corpus~\cite{panayotov2015librispeech}. We collect noise data from multiple datasets, including MUSAN~\cite{snyder2015musan}, RSG-10~\cite{steeneken1988description}, QUT-NOISE~\cite{dean2010qut}, the Environmental Noise dataset~\cite{saki2016smartphone}, UrbanSound~\cite{Urban}, the Nonspeech dataset~\cite{hu2010tandem}, and colored noise~\cite{deepmmse}. Noise recordings longer than 30 seconds are split into clips of up to 30 seconds, resulting in $6\,809$ noise clips. For validation, $1\,000$ speech utterances and noise clips are randomly drawn (without replacement) and mixed at SNR levels randomly sampled from -10 to 20 dB in 1 dB intervals, which generates $1\,000$ noisy-clean pairs. Four real-world noise sources are extracted for testing: \textit{F16}, \textit{factory welding}, and \textit{voice babble} from the RSG-10 dataset, and \textit{street music} from the UrbanSound dataset. For each of the four noise sources, twenty clean speech clips are randomly drawn from Librispeech \textit{test-clean-100} corpus~\cite{panayotov2015librispeech} and mixed with a random clip of the noise source at five SNR levels: $\left\{ -5 \text{ dB}, 0 \text{ dB}, 5 \text{ dB}, 10 \text{ dB}, 15 \text{ dB} \right\}$. This generates $400$ noisy mixtures for evaluation. All audio signals are sampled at 16 kHz. The STFT magnitude spectrum is computed using a square-root Hann window of 32 ms (512 samples) with a 16 ms frameshift (256 samples).} 

\subsection{Implementation Details}

\textcolor{black}{The configuration details for all the backbones are provided here. The model name is represented in the format ``backbone network--number of building blocks''. Transformer and Conformer backbones respectively consist of $N\!=\!4$ stacked Transformer and Conformer blocks in both causal and noncausal configurations, with the parameters following~\cite{10446337,ripple}, the number of attention heads $H\!=\!8$, model dimension $d_{model}\!=\!256$, and $d_{f\!f}\!=1\!024$. In addition, both absolute (Sinusoidal PE, SinPE) and relative (RoPE) PEs are employed to further boost the Transformer backbone in a noncausal configuration~\cite{10446337} for a systematic and fair comparison. For the causal configuration, xLSTM and Mamba backbones with both $5$ and $7$ stacked building blocks are designed for comparison with Transformer. xLSTM and Mamba backbones with 14 and 13 stacked blocks, respectively, are designed for comparison with Conformer. For the noncausal configuration, BixLSTM (Cascaded and Parallel) and BiMamba with both $3$ and $4$ blocks are designed for comparison with Transformer, and BixLSTM and BiMamba with $7$ blocks are designed for comparison with Conformer. We employ the default parameter configurations for Mamba~\cite{mamba} and xLSTM~\cite{beck2024xlstm} blocks.}

\begin{table}[!htbp]
    \centering
    \scriptsize
    \small
    \def\arraystretch{1.1}
    \setlength{\tabcolsep}{5.0pt}
    \setlength{\abovetopsep}{0pt}
    \setlength\belowbottomsep{0pt} 
    \setlength\aboverulesep{0pt} 
    \setlength\belowrulesep{0pt}
    \caption{\textcolor{black}{The comparison results of WB-PESQ scores for the models in both causal and non-causal configurations. The symbols ``\ding{52}'' and ``\ding{55}'' denote the causality and non-causality, respectively. ``\#Para.'' and ``Cau.'' denote the number of parameters and whether the model is causal.}}
    \label{tab1}
    \vspace*{0.05in}
    \begin{tabular}{@{}llcccccc@{}}
        \toprule[1.0pt]
         & & & \multicolumn{5}{c}{\textbf{Input SNRs}}\\  
        \cline{4-8}
        \textbf{Model} & \textbf{\#Para.} & \textbf{Cau.} & -5 & 0 & 5 & 10 & 15 \\  
        \hline
        \textcolor{black}{Noisy} & -- & -- & 1.05 & 1.07 & 1.13 & 1.31 & 1.64 \\
        \textcolor{black}{Transformer-4} & 3.29M & \ding{52} & 1.19 & 1.43 & 1.79 & 2.26 & 2.75  \\
        \textcolor{black}{Mamba-5} & 2.32M & \ding{52} & 1.20 & 1.45 & 1.85 & 2.33 & 2.81  \\
        \textcolor{black}{xLSTM-5}  & 2.21M & \ding{52} & 1.21 & 1.46 & 1.84 & 2.33 & 2.82  \\
        \textcolor{black}{Mamba-7}  & 3.20M & \ding{52} & 1.21 & 1.47 & \textbf{1.90} & \textbf{2.37} & 2.83  \\
        \textcolor{black}{xLSTM-7}  & 3.04M & \ding{52} & \textbf{1.22} & \textbf{1.48} & 1.88 & \textbf{2.37} & \textbf{2.84} \\
                
        \hline
        \textcolor{black}{Conformer-4} & 6.22M & \ding{52} & 1.24 & 1.51 & 1.93 & 2.41 & 2.88 \\
        \textcolor{black}{xLSTM-14} & 5.95M & \ding{52} & \textbf{1.25} & \textbf{1.53} & \textbf{1.96} & \textbf{2.44} & \textbf{2.91}  \\
        \textcolor{black}{Mamba-13} & 5.83M & \ding{52} & \textbf{1.25} & 1.52 & \textbf{1.96} & 2.43 & 2.90  \\
        \hline
        \hline
        \textcolor{black}{Transformer-4} & 3.29M & \ding{55} & 1.20 & 1.44  & 1.81 & 2.26 & 2.73    \\
        \, \textcolor{black}{+SinPE}    & 3.29M & \ding{55} & 1.26 & 1.56  & 1.99 & 2.48 & 2.94    \\
        \, \textcolor{black}{+RoPE}      & 3.29M & \ding{55} & 1.28 & 1.57  & 1.99 & 2.49 & 2.95    \\
        \textcolor{black}{C-BixLSTM-3}   & 2.63M & \ding{55} & 1.29 & 1.61  & 2.05 & 2.53 & 2.97    \\
        \textcolor{black}{P-BixLSTM-3}   & 2.63M & \ding{55} & 1.26 & 1.56  & 2.01 & 2.51 & 2.98    \\
        \textcolor{black}{C-BixLSTM-4}   & 3.46M & \ding{55} & \textbf{1.32} & \textbf{1.66} & \textbf{2.13} & \textbf{2.61} & \textbf{3.05}  \\
        \textcolor{black}{P-BixLSTM-4}   & 3.46M & \ding{55} & 1.29 & 1.62 & 2.07 & 2.55 & 3.01  \\
        \textcolor{black}{BiMamba-3}     & 2.76M & \ding{55} & 1.26 & 1.57 & 2.02 & 2.53 & 2.98  \\
        \textcolor{black}{BiMamba-4}     & 3.64M & \ding{55} & 1.30 & 1.63 & 2.09 & 2.58 & 3.03  \\
        \hline
        \textcolor{black}{Conformer-4} & 6.22M & \ding{55} & 1.35 & 1.70 & 2.17 & 2.66 & 3.09  \\
        \, \textcolor{black}{+SinPE}  & 6.22M & \ding{55} & 1.35 & 1.70 & 2.16 & 2.67 & 3.12  \\
        \, \textcolor{black}{+RoPE}    & 6.22M & \ding{55} & \textbf{1.36} & 1.71 & 2.17 & 2.67 & 3.11  \\
        \textcolor{black}{C-BixLSTM-7} & 5.95M & \ding{55} & 1.35 & \textbf{1.72} & 2.20 & 2.69 & 3.13  \\
        \textcolor{black}{P-BixLSTM-7} & 5.95M & \ding{55} & 1.35 & \textbf{1.72} & \textbf{2.21} & \textbf{2.70} & 3.13  \\
        \textcolor{black}{BiMamba-7}   & 6.26M & \ding{55} & 1.35 & \textbf{1.72} & \textbf{2.21} & \textbf{2.70} & \textbf{3.14}  \\
        
        
        \toprule[1.0pt]
    \end{tabular}
\end{table}

\begin{table}[!h]
    \centering
    \scriptsize
    \small
    \def\arraystretch{1.15}
    \setlength{\tabcolsep}{4.0pt}
    \setlength{\abovetopsep}{0pt}
    \setlength\belowbottomsep{0pt} 
    \setlength\aboverulesep{0pt} 
    \setlength\belowrulesep{0pt}
    \caption{\textcolor{black}{The comparison results of ESTOI (in \%) scores for the models in both causal and non-causal configurations.}}
    \label{tab2}
    \vspace*{0.05in}
    \begin{tabular}{@{}llcccccc@{}}
        \toprule[1.0pt]
         & & & \multicolumn{5}{c}{\textbf{Input SNRs}}\\  
        \cline{4-8}
        \textbf{Model} & \textbf{\#Para}. & \textbf{Cau.} & -5 & 0 & 5 & 10 & 15 \\  
        \hline
        \textcolor{black}{Noisy} & -- & -- & 27.91 & 42.14 & 57.21 & 71.11 & 82.22 \\
        \textcolor{black}{Transformer-4} & 3.29M & \ding{52} & 43.46 & 61.18 & 70.78 & 84.27 & 90.17  \\
        \textcolor{black}{Mamba-5}  & 2.32M & \ding{52} & 44.90 & 62.61 & 76.16 & 85.05 & 90.66 \\
        \textcolor{black}{xLSTM-5}  & 2.21M & \ding{52} & 44.85 & 62.58 & 76.10 & 84.99 & 90.70  \\
        \textcolor{black}{Mamba-7}  & 3.20M & \ding{52} & \textbf{46.37} & 63.71 & 76.81 & 85.34 & 90.82   \\
        \textcolor{black}{xLSTM-7}  & 3.04M & \ding{52} & 46.18 & \textbf{63.87} & \textbf{76.87 }& \textbf{85.46} & \textbf{90.84}  \\
                
        \hline
        \textcolor{black}{Conformer-4} & 6.22M & \ding{52} & 47.93 & 64.87 & 77.54 & 85.83 & 91.05  \\
        \textcolor{black}{xLSTM-14} & 5.95M & \ding{52} & \textbf{48.54} & \textbf{65.50} & \textbf{77.95} & \textbf{86.02} & 91.23  \\
        \textcolor{black}{Mamba-13} & 5.83M & \ding{52} & 48.20 & 65.16 & 77.81 & 86.00 & \textbf{91.25}   \\
        \hline
        \hline
        \textcolor{black}{Transformer-4} & 3.29M & \ding{55} & 42.31 & 60.44 & 74.76 & 84.31 & 90.24  \\
        \, \textcolor{black}{+SinPE}    & 3.29M & \ding{55} & 47.85 & 65.03 & 77.76 & 86.00 & 91.22   \\
        \, \textcolor{black}{+RoPE}      & 3.29M & \ding{55} & 48.13 & 65.21 & 77.93 & 86.13 & 91.34 \\
        \textcolor{black}{C-BixLSTM-3}   & 2.63M & \ding{55} & 49.86 & 66.60 & 78.71 & 86.51 & 91.45   \\
        \textcolor{black}{P-BixLSTM-3}   & 2.63M & \ding{55} & 47.94 & 65.22 & 78.02 & 86.23 & 91.35 \\
        \textcolor{black}{C-BixLSTM-4}   & 3.46M & \ding{55} & \textbf{51.93} & \textbf{68.30} & \textbf{79.81} & \textbf{87.11} & \textbf{91.81}   \\
        \textcolor{black}{P-BixLSTM-4}   & 3.46M & \ding{55} & 50.25 & 67.10 & 79.18 & 86.77 & 91.66   \\
        \textcolor{black}{BiMamba-3}     & 2.76M & \ding{55} & 48.62 & 66.12 & 78.44 & 86.43 & 91.47   \\
        \textcolor{black}{BiMamba-4}     & 3.64M & \ding{55} & 49.89 & 67.05 & 79.20 & 86.79 & 91.69   \\
        \hline
        \textcolor{black}{Conformer-4} & 6.22M & \ding{55} & 53.43 & 69.59 & 80.47 & 87.55 & 92.09  \\
        \, \textcolor{black}{+SinPE}  & 6.22M & \ding{55} & 53.56 & 69.81 & 80.57 & 87.57 & 92.11  \\
        \, \textcolor{black}{+RoPE}    & 6.22M & \ding{55} & 53.65 & 69.71 & 80.50 & 87.59 & \textbf{92.12}  \\
        \textcolor{black}{C-BixLSTM-7} & 5.95M & \ding{55} & \textbf{54.42} & \textbf{70.00} & \textbf{80.70} & \textbf{87.67} & \textbf{92.12}  \\
        \textcolor{black}{P-BixLSTM-7} & 5.95M & \ding{55} & 53.41 & 69.37 & 80.37 & 87.46 & 92.03  \\
        \textcolor{black}{BiMamba-7}   & 6.26M & \ding{55} & 53.72 & 69.81 & 80.66 & 87.63 & 92.09  \\
        
        
        \toprule[1.0pt]
    \end{tabular}
\end{table}

\begin{table}[!h]
    \centering
    \scriptsize
    \small
    \def\arraystretch{1.15}
    \setlength{\tabcolsep}{7.5pt}
    \setlength{\abovetopsep}{0pt}
    \setlength\belowbottomsep{0pt} 
    \setlength\aboverulesep{0pt} 
    \setlength\belowrulesep{0pt}
    \caption{\textcolor{black}{The comparison results of three composite metrics (CSIG, CBAK, and COVL) for the models in both causal and non-causal configurations.}}
    \label{tab3}
    \vspace*{0.1in}
    \begin{tabular}{@{}llcccc@{}}
        \toprule[1.0pt]
        \textbf{Model} & \textbf{\#Para.} & \textbf{Cau.} & \textbf{COVL} & \textbf{CSIG} & \textbf{CBAK}  \\  
        \hline
        \textcolor{black}{Noisy} & -- & -- & 1.67 & 2.26 & 1.80 \\
        \textcolor{black}{Transformer-4} & 3.29M & \ding{52} & 2.48 & 3.18 & 2.53   \\
        \textcolor{black}{Mamba-5}  & 2.32M & \ding{52} & 2.53 & 3.22 & 2.59  \\
        \textcolor{black}{xLSTM-5}  & 2.21M & \ding{52} & 2.52 & 3.21 & 2.57  \\
        \textcolor{black}{Mamba-7}  & 3.20M & \ding{52} & 2.56 & 3.25 & 2.61  \\
        \textcolor{black}{xLSTM-7}  & 3.04M & \ding{52} & 2.56 & 3.26 & 2.61  \\
                
        \hline
        \textcolor{black}{Conformer-4} & 6.22M & \ding{52} & 2.61 & 3.32 & 2.63 \\
        \textcolor{black}{xLSTM-14} & 5.95M & \ding{52} & 2.63 & 3.33 & 2.67 \\
        \textcolor{black}{Mamba-13} & 5.83M & \ding{52} & 2.63 & 3.34 & 2.66 \\
        \hline
        \hline
        \textcolor{black}{Transformer-4} & 3.29M & \ding{55} & 2.47 & 3.17 & 2.55  \\
        \,\textcolor{black}{+SinPE}     & 3.29M & \ding{55} & 2.64 & 3.33 & 2.66  \\
        \,\textcolor{black}{+RoPE}       & 3.29M & \ding{55} & 2.65 & 3.33 & 2.67  \\
        \textcolor{black}{C-BixLSTM-3}   & 2.63M & \ding{55} & 2.70 & 3.40 & 2.68  \\
        \textcolor{black}{P-BixLSTM-3}   & 2.63M & \ding{55} & 2.66 & 3.35 & 2.67  \\
        \textcolor{black}{C-BixLSTM-4}   & 3.46M & \ding{55} & \textbf{2.77} & \textbf{3.48} & \textbf{2.76}  \\
        \textcolor{black}{P-BixLSTM-4}   & 3.46M & \ding{55} & 2.72 & 3.43 & 2.72  \\
        \textcolor{black}{BiMamba-3}     & 2.76M & \ding{55} & 2.67 & 3.36 & 2.66  \\
        \textcolor{black}{BiMamba-4}     & 3.64M & \ding{55} & 2.74 & 3.44 & 2.73  \\
        \hline
        \textcolor{black}{Conformer-4} & 6.22M & \ding{55} & 2.79 & 3.46 & 2.79  \\
        \, \textcolor{black}{+SinPE}  & 6.22M & \ding{55} & 2.80 & 3.47 & 2.80  \\
        \, \textcolor{black}{+RoPE}    & 6.22M & \ding{55} & 2.81 & 3.50 & 2.80  \\
        \textcolor{black}{C-BixLSTM-7} & 5.95M & \ding{55} & \textbf{2.85} & \textbf{3.56} & \textbf{2.83}  \\
        \textcolor{black}{P-BixLSTM-7} & 5.95M & \ding{55} & \textbf{2.85} & 3.55 & \textbf{2.83}  \\
        \textcolor{black}{BiMamba-7}   & 6.26M & \ding{55} & \textbf{2.85} & 3.55 & \textbf{2.83}  \\
        
        \toprule[1.0pt]
    \end{tabular}
\end{table}

\textbf{Training Methodology}. \textcolor{black}{We use a mini-batch of 10 speech clips for each training step. For each mini-batch, the noisy mixture clips are dynamically generated at training time. Each clean speech clip is degraded by a random segment of randomly selected noise recording, at an SNR randomly chosen from $-10$ dB to $20$ dB in $1$ dB steps. A total of $150$ epochs are used to train all the models, where the mean-square error (MSE) \textcolor{black}{in frequency domain} is used as the objective function for mask estimation. We employ the Adam optimizer gradient learning, adopting the hyper-parameters from~\cite{transformer}, $\beta_{1}\!=\!0.9$, $\beta_{2}\!=\!0.98$, and $\epsilon\!=\!1\text{e}{-9}$. Following~\cite{10446337,transformer}, a warm-up scheduler is employed for tuning the learning rate during training for all models. The scheduler is defined as: $lr =\textrm{min} \left(step\_n^{-0.5}, step\_n \cdot step\_w^{-1.5}\right)\cdot d_{model}^{-0.5}$, with $step\_n$ and $step\_w$ denoting the number of training steps and warm-up stage steps, respectively. $step\_w$ is set to $40\,000$ as~\cite{10446337,tfaj}. Gradients are clipped to between $-1$ and $1$ using the gradient clipping method. \textcolor{black}{The same random seed is used for model initialization and data pipeline across models.} All experiments are run on an NVIDIA V100-PCIE-40GB graphics processing unit.}

\subsection{Results and Analysis}

\textcolor{black}{We assess the quality and intelligibility of speech signals using perceptual evaluation of speech quality (PESQ)~\cite{pesq} and extended short-time objective intelligibility (ESTOI)~\cite{jensen2016algorithm}. Three composite metrics, CSIG, CBAK, and COVL~\cite{hu2007evaluation}, are used as predictors of the mean opinion score (MOS) for signal distortion, background noise intrusiveness, and overall signal quality.}

\textcolor{black}{Tables~\ref{tab1}-\ref{tab2} present the evaluation scores attained by models in PESQ and ESTOI, respectively. For the causal configuration, overall, Mamba and xLSTM backbones outperform the Transformer and Conformer across five SNRs while maintaining lower parameter overheads. Taking the case of 5 dB SNR, for example, compared to the causal Transformer-4 (3.29M), Mamba-7 (3.20M) and xLSTM-7 (3.04M) achieve PESQ gains of 0.11 and 0.09, and ESTOI gains of 6.03\% and 6.09\%, respectively. Mamba-5 (2.32M) and xLSTM-5 (2.21M) still exhibit superior performance to Transformer-4. Similarly, Mamba-13 (5.83M) and xLSTM-14 (5.95M) backbones are also observed to slightly outperform the Conformer-4 (6.22M). In comparison to Mamba, xLSTM exhibits comparable or slightly better performance. In the noncausal configuration, similarly, BixLSTM and BiMamba achieve superior performance than Transformers (with and without position embeddings). It also can be seen that in line with recent findings~\cite{10446337}, noncausal Transformer significantly benefits from the use of position embeddings. In the 10 dB SNR case, compared to Transformer-4 with RoPE (3.29M), cascaded (C) BixLSTM-4 (3.46M) and BiMamba-4 (3.64M) provide 0.12 and 0.09 PESQ gains, and 0.98\% and 0.66\% ESTOI gains. With smaller model sizes, C-BixLSTM-3 (2.63M) and BiMamba-4 (2.76M) also show superior performance. C-BixLSTM and BiMamba exhibit comparable or superior performance than Conformers. Overall, among BiMamba, C-BixLSTM, and P-BixLSTM, C-BixLSTM performs slightly better. Table~\ref{tab3} reports the comparison results in CSIG, CBAK, and COVL scores. A similar performance trend to that shown in Tables~\ref{tab1}-\ref{tab2} can be observed. }

\begin{table}[!t]
    \centering
    \small
    \def\arraystretch{1.1}
    \setlength{\tabcolsep}{6.0pt}
    \setlength{\abovetopsep}{0pt}
    \setlength\belowbottomsep{0pt} 
    \setlength\aboverulesep{0pt} 
    \setlength\belowrulesep{0pt}
    \caption{\textcolor{black}{The comparison results of training speed (seconds/step, $\times 10^{-2}$) and inference speed (real-time factor, RTF, $\times 10^{-4}$) across different test lengths (10s, 20s, and 40s).}}
    \label{tab4}
    \vspace*{0.05in}
    \begin{tabular}{@{}lcccccc@{}}
        \toprule[1.0pt]
        \multirow{2}{*}{\textbf{Model}} & \multirow{2}{*}{\textbf{\#Para.}} & \multirow{2}{*}{\makecell{\textbf{sec/step} \\ \textbf{($\times 10^{-2}$)}}} &\multicolumn{3}{c}{\textbf{RTF} (\textbf{$\times 10^{-4}$})} \\
        \cline{4-6}
        & & & 10s & 20s & 40s \\  
        \hline
        \textcolor{black}{Transformer-4}   & 3.29M & 6.91  &\textbf{1.112} & 1.537 & 2.642  \\
        \,\,\textcolor{black}{+SinAPE}     & 3.29M & 6.94  &1.116 & 1.539 & 2.643  \\
        \,\,\textcolor{black}{+RoPE}       & 3.29M & 7.67  &1.534 & 1.827 & 2.987  \\
        \textcolor{black}{C-BixLSTM-4}     & 3.46M & 25.67 &3.035 & 4.033 & 6.148  \\
        \textcolor{black}{P-BixLSTM-4}     & 3.46M & 25.64 &3.041 & 4.022 & 6.151  \\
        \textcolor{black}{BiMamba-4}       & 3.64M & \textbf{5.66}  &1.212 & \textbf{0.707} & \textbf{0.664}  \\
        \hline
        \hline
        
        \textcolor{black}{Conformer-4}    & 6.22M & 9.26 & \textbf{1.569} & 1.959 & 3.020  \\
        \,\, \textcolor{black}{+SinAPE}   & 6.22M & 9.26 & 1.582 & 1.963 & 3.023  \\
        \,\, \textcolor{black}{+RoPE}     & 6.22M & 9.99 & 2.014 & 2.242 & 3.363  \\
        \textcolor{black}{C-BixLSTM-7}    & 5.95M & 44.81 & 5.254 & 7.052 & 10.70  \\
        \textcolor{black}{P-BixLSTM-7}    & 5.95M & 44.48 & 5.264 & 7.053 & 10.74  \\
        \textcolor{black}{BiMamba-7}      & 6.26M & \textbf{7.12} & 2.072 & \textbf{1.195} & \textbf{1.139}  \\
        
        \toprule[1.0pt]
    \end{tabular}
\end{table}
\textcolor{black}{\textbf{Training \& Inference speeds.} In Table~\ref{tab4}, we compare the training and inference speeds of the models, measured in terms of training time per step (seconds/step) and real-time factor (RTF). The RTF is computed as the ratio of the processing time of a speech utterance to its duration~\cite{cleanunet}. The RTF values were measured on an NVIDIA Tesla A100 GPU, averaged over 20 runs, and are evaluated across different test lengths (10s, 20s, and 40s), using a batch size of 4 noisy mixtures~\cite{cleanunet,tfaj}. It can be observed that BiMamba exhibits a faster training speed than Transformer, Conformer, and BixLSTM. For the 10s test length case, Transformer and Conformer afford a slightly better inference speed than BiMamba. As the test length increases, BiMamba demonstrates a clear superiority over Transformer and Conformer in inference speed. For the 20s and 40s test length cases, Conformer-4 respectively shows approximately $2$ and $3$ times the RTF of BiMamba-4.}

\section{Conclusion}\label{sec:5}
\textcolor{black}{In this paper, we investigated the effectiveness of the recently introduced xLSTM for speech enhancement and conducted a systematic and fair comparative study of advanced long-context modeling backbone networks, including Transformer, Conformer, Mamba, and xLSTM. Our comprehensive experimental results demonstrate that, overall, (Bi)xLSTM and (Bi)Mamba exhibit superior performance compared to self-attention-based Transformer and Conformer, in causal and non-causal configurations. xLSTM achieves comparable or superior performance compared to Mamba but suffers from slower training and inference speeds.  In contrast, Mamba demonstrates a clear advantage in efficiency, particularly for long input speech.}

\section{Acknowledgment}\label{sec:ack}

This research work is supported in part by Shenzhen Research Institute of Big Data under Grant No. K00120240007, Shenzhen Science and Technology Program (Shenzhen Key Laboratory Grant No. ZDSYS20230626091302006), and the Program for Guangdong Introducing Innovative and Enterpreneurial Teams, Grant No. 2023ZT10X044.




\clearpage
\bibliographystyle{IEEEtran}
\bibliography{refs25}







\end{document}